# NCARD: Improving Neighborhood Construction by Apollonius Region Algorithm based on Density

Shahin Pourbahrami[1], Leyli Mohammad Khanli[*,1], Sohrab Azimpour[2]

Computer Engineering Department, Faculty of Electrical and Computer Engineering, University of Tabriz, Tabriz, Iran[1]
Mathematic Department, Faculty of Mathematic, University of Farhangian, Tehran, Iran[2]

**Abstract**
Due to the increased rate of information in the present era, local identification of similar and related data points by using neighborhood construction algorithms is highly significant for processing information in various sciences. Geometric methods are especially useful for their accuracy in locating highly similar neighborhood points using efficient geometric structures. Geometric methods should be examined for each individual point in neighborhood data set so that similar groups would be formed. Those algorithms are not highly accurate for high dimension of data. Due to the important challenges in data point analysis, we have used geometric method in which the Apollonius circle is used to achieve high local accuracy with high dimension data. In this paper, we propose a neighborhood construction algorithm, namely Neighborhood Construction by Apollonius Region Density (NCARD). In this study, the neighbors of data points are determined using not only the geometric structures, but also the density information. Apollonius circle, one of the state-of-the-art proximity geometry methods, Apollonius circle, is used for this purpose. For efficient clustering, our algorithm works better with high dimension of data than the previous methods; it is also able to identify the local outlier data. We have no prior information about the data in the proposed algorithm. Moreover, after locating similar data points with Apollonius circle, we will extract density and relationship among the points, and a unique and accurate neighborhood is created in this way. The proposed algorithm is more accurate than the state-of-the-art and well-known algorithms up to almost 8-13% in real and artificial data sets.

## 1. Introduction

With mass production of electronic data in the world today, we encounter the problem of data processing and savin. Therefore, data mining is a scientific and applied method for extracting a valid relationship between data patterns to process and save the data in the data set. Data mining helps model systems, explore decision region, and organize the data. The main purpose behind data mining is clustering and classifying the data as well as exploring knowledge. Data mining has a broad range of applications in clustering including intrusion detection (Al-Jarrah et al., 2016), document clustering (Pei et al., 2014, Gu et al 2013), face recognition (Cao et al., 2015, Pei et al., 2015) and tumor clustering from cancer gene expression (Yu et al 2015, Yu et al 2014). The aim of the present paper is examining data clustering based on identifying neighborhood points in the data. In data mining, neighborhood is defined as finding a set of similar data on a local basis in such a way that there is highest similarity inside the newly found group and minimal similarity among the groups (İnkaya et al., 2015). The aim of constructing data sets is analysing them in different scientific areas in order to extract the significant information out of them. In other words, the expression 'Data Mining' means extracting the implicit information or sepecific patterns and relationships which are needed in a large amount of data in one or more big databases.

In order to improve neighborhood construction by the use of geometric structures and neighborhood topology, algorithms have been suggested. For example, NC (Neighborhood Construction) and Gabriel's graph neighbors are two geometric algorithms utilized for examining direct/indirect relationships among the points (İnkaya et al., 2015, Gabriel & Sokal, 1969, Guedes et al., 2016, Hashem et al., 2015, Koontz et al., 1976, Urquhart, 1982). Geometric methods have the advantage of high accuracy in finding neighborhood points with high similarity by using highly efficient geometric structures. For each given point in neighborhood data set, the

Corresponding Author. *
*Email Addresses:* sh.pourbahrami@tabrizu.ac.ir (Shahin Pourbahrami), l-khanli@tabrizu.ac.ir (Leyli Mohammad Khanli), azimpour@cfu.ac.ir (Sohrab Azimpour).

available geometric methods should be carefully examined to form similar groups. NC and Gabriel's graph suffer from the disadvantage of failing to have high accuracy with large scale data.

k-Nearest Neighbor (kNN) is a machine learning algorithm that has been studied in pattern recognition method for several decades (Duda et al., 2012, Jarvis & Patric 1973, Lu & Fu 1978, Qin et al., 2018, Tao et al., 2006). This algorithm is among the most commonly used methods in identifying neighborhood points by using density and graph base models.

In DBSCAN algorithm (Ester et al. 1996), epsilon radius (ε) of the circle is used to define neighborhood epsilon (Wang et al., 2018). In a circle with a radius of epsilon, if the points inside the circle are defined at a distance of at least MinPts from the concerned point, they are called as core neighborhood points for that point. However, if the points are located out of the defined points, they are labelled as border or noise points. Decision region and the density among these points in neighborhood space relates them in clustering. One of the shortcomings of this algorithm is definition of epsilon parameter (ε). Ertoz et al. utilized neighborhood by graph based kNN on the categorical data (Ertöz et al 2003). They managed to increase accuracy, but the number of parameters for regulations in the algorithm increased to ε, k, and MinPts. Based on the mentioned algorithms, the local features are useful for extracting similarities on clustering and classification in data mining algorithms.

Neighborhood Construction by Apollonius Region (NCAR) algorithm initially finds high density points known as target points. For each target point, NCAR algorithm finds the farthest distant point, and according to the target points and distant points, the center and radius of Appollonius circles are determined leading to the formation of primary Apollonius circles (Pourbahrami et al., 2018). However, this algorithm is not highly accurate for large amounts and high dimension of data. In the proposed algorithm, in order to remove the shortcomings of NCAR algorithm, the density and internal relationships inside the circles were used to promote the accuracy of the algorithm. For each single point around the target points, Appollonius circle is drawn and the neighborhood region is defined.

In NCARD algorithm, geometric Appollonius method was used in order to find the exact neighborhood region between the points. This algorithm works better with large data sets than the previous methods; it is also able to identify the local outlier data. The structure of this algorithm is based on the geometry and distance between the points. In this way, Apollonius helps offer a new and accurate definition of neighborhood. Moreover, by extracting the density and the relationship among the points, a unique and accurate neighborhood is created. In other words, the density between the points inside Appolnius circle is examined in order to increase the accuracy between neighborhood points. This algorithm does not require a knowledge of the data, and it is used in clustering. To sum up, the contributions of our study are fourfold: (1) the neighbors of a data point are determined using not only the geometric structures, but also the density and decision reigens information. (2) The neighborhood of target points are uniquely determined. (3) It is assumed that there is no prior information about the data sets. (4) This algorithm identifies the outliers using geometric structure and decision regions.

In the rest of this paper, the following section will cover a review of the studies conducted on identifying neighborhood. In section three, the proposed algorithm and the stages in creating it will be elaborated upon. Section four will focus on evaluation criteria for examining the performance of the proposed algorithm. Later, in section five the results of the experiments on different data sets using our proposed method will be presented with a comparative examination of the results obtained with other algorithms. Finally, the last secticon or conclusion and suggestions for further studies will be presented.

## 2. Review of Related Literature

It is indispensable that we need to find data point neighborhood and connection for clustering domains. The reason for creating neighborhood construction algorithms is exploring the decision region among the data for the purpose of analyzing and processing them, as well as their consequent use in different scientific areas.

Neighborhood construction with Gabriel's graph in NC algorithm is geometry-based and density between the points (İnkaya et al., 2015). In this line, first a given point is chosen, and all other points are arranged according to it (the nearest point appears first, and the most distant points come later). In the second step, by means of Gabriel's graph, the direct/indirect connections are obtained and their densities in a single set are measured. In case there is a decrease in the density inside the data sets or the points in the data set, a new neighborhood group forms. In the next step, the neighborhood constructed in the intersection of break points and the neighborhood group of a particular point is carefully examined. By definition, two points are called neighbors if an intersection is formed among the obtained neighborhood points. However, when there is no intersection, it is most likely that independent and separate groups will be formed. In the fourth step, the mutual relationships among nearest neighbors are examined so that the outlier data would be identified. The shortcoming of this algorithm is its high complex. As such, it is more suitable for examining the data sets in small databases.

In computational geometry and graph theory, *β*-skeleton is the skeleton of geometry-based algorithm which is defined on a set of angle between geometric points. In *β*-skeleton algorithm, in order to determine neighborhood region, the value of angle (*Θ*) should be identified. In the right angle of $90^0$, this algorithm practices like Gabriel's algorithm (Langerman et al., 2009, Toussaint et al., 1988, Yang et al., 2016).

For two points on a plate, if a Voronoi edge is shared between them, then those two points would be considered as neighbours (Liu et al., 2 2008). This type of neighborhood is the base for making Delaunay Triangulation (DT) in which each edge of the triangle shows a neighborhood structure between the two points. One area is allotted for each point of the data, and they are termed as Voronoi cells. For each data point out of the whole points of the defined area, all edges of the area are close to the point creating that area. One of the uses of Voronoi diagram is Delaunay triangulation for clustering the points. In computational geometry, Delaunay triangulation is used to refer to a set of points labelled *P* in a plate called *R*. One triangulation is called as DT (P) in a way that none of the *P* points in triangulation is located inside circumscribed circles of the triangles of DT (P). In other words, let's assume that we have a finite set of points in a Euclidean space, and for each point we consider an area. The areas include points in a way that the distance of each point inside the area from a specific point is less than or equal to the distance of the other edge with the respective point. Two points are located in the same cluster if they are adjacent to the convex hull and have at least one common edge. In AMOE approach, standard deviation and mean edge length are used to identify the clusters based on local and overall Delaunay triangulation edge (Estivill & Lee, 2000). This approach uses no parameter and graph base, but it is designed for data sets in two dimensions, and it has no use for higher dimensions.

In kNN algorithm among the points in a data set, some points are located using a similarity criterion parameter k which is used for predicting data points in each selected set (Jarvis & Patric 1973, Lu & Fu 1978, Qin et al., 2018, Tao et al., 2006). This algorithm considers the data points as belonging to a cluster having greatest number of votes among the k of the nearest neighbor or mutual neighborhoods. Despite its advantages, this algorithm suffers from the two following problems: The first problem with this algorithm concerns the determination of k value by the user which is highly important for extracting decision region and local features among the points. The second problem lies in the fact that in determining neighborhood, it is also highly important to decide on the criterion of similarity among the points. Determining k value is a very important and key issue in this algorithm, and wrong estimation of k value leads to error in defining neighborhood. kNN algorithm is a method for locating mutual decision region among the points, and the equation between them is defined based on similarity criterion (Abbas & Shoukry, 2012, Brito et al., 1997, Hu & Bhatnagar,. 2012, Sardana & Bhatnagar, 2014). These algorithms are more often used in graph base methods for identification of neighborhood as well as mutual neighborhood among the data points in different data sets.

In graph base neighborhood algorithms, the local relationships among the observations can be extracted effectively. DENGP is a graph base algorithm (Yu & Kim 2016a, Yu & Kim 2016b). This algorithm shows all data points based on Mutual k-nearest graph in the data space. The density of each node is calculated to determine high density areas (calculation of density coefficients for all nodes). All nodes are divided into two groups: core nodes (high density nodes) and surrounding nodes (low density nodes). In the next step, all surrounding nodes are temporarily removed from the data points, and core nodes are divided into sub-clusters (these subgroups are agglomerated to maximum internal connection in the clusters and are integrated into each other hierarchically). Finally, the surrounding nodes are added to the clusters created with core nodes based on weighed majority voting, and the accuracy of node disposal is examined to determine whether the nodes are connected to their own respective clusters or not. The aim of this algorithm is to minimize the distances inside the sets and to maximize the differences among the clusters.

The idea of Density Peaks Clustering (DPC) implies that cluster centers are denoted by a higher density than their neighbors and a relatively large distance from data points with higher densities (Li & Tang, 2018, Rodriguez & Laio, 2014). It is based on the approach that cluster centers are obtained with high densities of their neighbors with a rather high distance from high density data points. For each sample, these quantities are calculated according to the matrix of data point distances. Unlike center-based algorithms, this algorithm does not need the parameter of the number of clusters and repetitive process. In DPC-KNN after recognition cluster centers, each data point is assigned to its nearest target (Rodriguez & Laio, 2014). DPC-PCA is a kind of algorithm which is formed according to principal component analysis which minimizes the data and allots 99% of the principal components to eigenvectors (Du et al., 2016). Observations of the comparisons between neighborhood construction algorithms in Table 1 indicate that most of the algorithms in geometry-based and density-based algorithms have many computations and are not suitable for high dimensions of the data sets.

**Table 1.** The advantages and disadvantages of neighborhood construction algorithms

| Neighborhood construction Algorithms | Advantages | Disadvantages |
|---|---|---|
| k-NN (Jarvis & Patric 1973, Lu & Fu 1978, Qin et al., 2018, Tao et al., 2006) | • Simple approach<br>• Nonparametric<br>• Intuitive approach<br>• Robust to outliers detection | • Determining neighborhood just according to the distance<br>• Sensitivity to irrelevant features<br>• Lack of theoretical basis |
| DBSCAN (Wang et al., 2018) | • Discover clusters of arbitrary shapes<br>• Outliers detection<br>• No need to define number of clusters | • Cannot find clusters effectively with different density<br>• Cannot work with several outliers<br>• Uncertainty of tuning the input parameters<br>• Not suitable for high-dimensional data<br>• There is not a unified and generally accepted approach to determine parameters |
| NC (İnkaya et al., 2015) | • Robust to outliers detection<br>• Being parameter-free<br>• Create unique set of neighbors<br>• Geometrically intuitive | • High computational cost<br>• Create fixed of number and neighbors region<br>• Suitable just for small data set<br>• Should be examined for each individual data point |
| $\beta$-skeleton (Langerman et al., 2009, Toussaint et al., 1988, Yang et al., 2016) | • Create changeable of neighbors region<br>• Geometrically intuitive | • Uncertainty of tuning the input parameters<br>• High computational cost<br>• Should be examined for each individual data point |
| DPC-KNN (Li & Tang, 2018, Rodriguez & Laio, 2014) | • Lack of repetition phase<br>• The centers are extracted from the basic features of the data points | • High computational cost<br>• Suitable just for small data set<br>• Need to define parameter<br>• The number of appropriate clusters obtained from the decision graph may not be equal and comparable with the number of ideal clusters.<br>• Not sensitive to the local geometric features |
| DPC-PCA (Du et al., 2016) | • Reducing dimensions of the data set<br>• The centers are extracted from the basic features of the data points | • High computational cost<br>• Suitable just for small data set<br>• Lack of any strategy in dealing with the outlier samples<br>• Would not yield an acceptable performance in overlapping data sets |
| NCAR (Pourbahrami et al., 2018) | • Robust to outliers detection<br>• Geometrically intuitive<br>• Low computational cost | • Suitable just for small data set |

## 3. Proposed Method

Geometric methods should be examined for each individual data point in neighborhood data set so that similar clusters would be formed. Those algorithms are not highly accurate for high dimension of data. Due to the significance of neighborhood construction on data points, our algorithm was presented with different data sets so that we could obtain high accuracy in constructing neighborhood with Apollonius circle. In other words, using distance and harmonious geometric Apollonius structure, the similarity among the data points in the data space is identified. Depending on data point neighborhood regions, the decision region among the points determines the density among the points so that the similarity among the points inside the groups is maximized while the similarity between the groups is minimized. Therefore, the goal is locating similar data groups in a set of data without any prior information about data clusters and the number of clusters.

### 3.1 Notation and definition

The notations that have been used throughout the paper to indicate the definitions and formula are presented Table 2.

**Table 2.** The notation of NCARD algorithm.

| | |
|---|---|
| M | Set of data points |
| $CP_m$ | Neighborhood sets of the Control Point |
| $T_m$ | Neighborhood sets of the Target point |
| T | Set of Target points (high density) |
| $F_P$ | Farthest data Point |
| $C_{AB}$ | Apollonius circle for points A and B |
| d(A, M) | Euclidean distance between points A and M |
| k | Distance ratio of d(A, M) over d(M, B) |
| DR (A, B, M) | Decision Region |
| D | Density sets of the data points |

**Definition of $C_{AB}$:** As Fig. 1 shows, Apollonius circle can be defined as the geometric position of the points in a Euclidean plane in which the distance of the given points from point A and B is determined to be k(k≠0). Eq. (1) (Pourbahrami et al., 2018, Hoshen, 1996, Partensky, 2008).

$$k = \frac{d(A,M)}{d(M,B)} \quad (1)$$

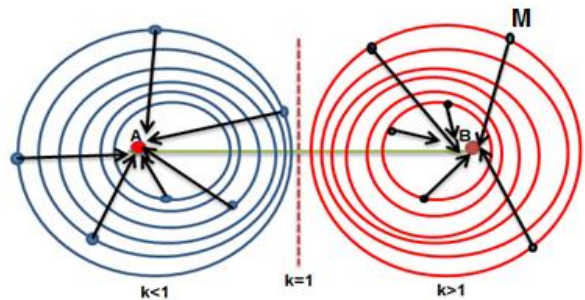

**Fig. 1.** Apollonius circles $C_{AB}$ (Pourbahrami et al., 2018)

**Definition of Decision Region (DR):** In Eq (2) with the condition of *k >1* and *k <1* decision regions are made.

$$DR(A,B,M) = \begin{cases} (C_{AB} \cap C(B,BM) \cap \text{upper or down plan of the line contain } AB) & \text{if } \dfrac{d(A,M)}{d(M,B)} \rangle 1 \\ (C_{AB} \cap C(A,AM) \cap \text{upper or down plan of the line contain } AB) & \text{if } \dfrac{d(A,M)}{d(M,B)} \langle 1 \end{cases} \quad (2)$$

**Definition of Density:** In Eq (3) with the condition of $k >1$, we can claim that *B* and *M* are directly related (density = 0) if the intersection of Apollonius circle with circle *C (B, BM)* and upper plane of the line containing *AB* is empty if *M* is in upper part as shown in Fig.2. Otherwise, it is said that *B* and *M* have indirect relationship (density = 2, number of members in decision region) in Fig.3. It means:

$$D(B,M) = \begin{cases} \text{direct} & \text{if } (C_{AB} \cap C(B,BM) \cap \text{upper plan of the line contain } AB) = \varnothing \\ \text{indirect} & \text{if } (C_{AB} \cap C(B,BM) \cap \text{upper plan of the line contain } AB) \neq \varnothing \end{cases} \quad (3)$$

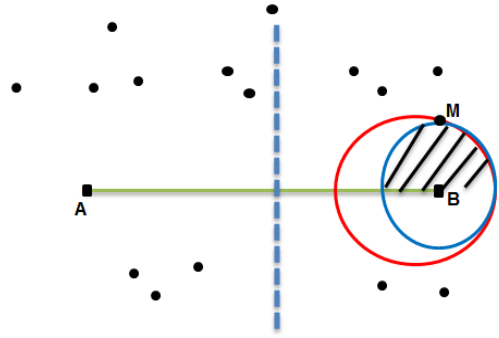

**Fig.2.** The direct connection between B and M where density is 0.

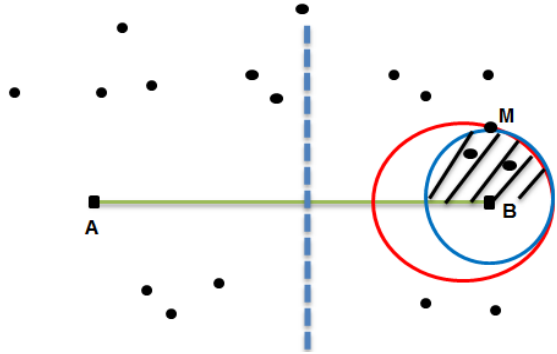

**Fig.3.** The indirect connection between B and M where density is 2.

## 3.2. Neighborhood Construction by Apollonius Region Density algorithm

   NCARD algorithm uses density and the decision region among the data points to determine neighborhood and explores the direct/indirect relationship among the data points using Apollonius circle. The goal is enhancing the accuracy of NCAR algorithm (Pourbahrami et al., 2018) in order to obtain different forms of clustering inside Apollonius circles by defining Apollonius region and density among the points. In case new clusters are formed inside the obtained circles (as a result of neighborhood with NCARD algorithm) we will try to explore them. The algorithm has three steps the first of which is NCAR algorithm which leads to the creation of Apollonius neighborhood circles. The next step is finding the decision region among the data points inside the circles. In the third step, density among the data points inside the circles and the target points above the circle are examined based on the direct/indirect relationships detected in the second step. In other words, in the first step, we identify the data points and, in subsequent steps, maximize the similarity of the data points in clusters.

**Step 1: Grouping data points based on NCAR algorithm**

The first step in our algorithm is extracting the original neighborhood groups with Apollonius circle which is included in NCAR algorithm (Pourbahrami et al., 2018). This step includes three parts the first of which involves detecting target points (T). The second part includes locating the farthest point ($F_P$) from the target points. Drawing Apollonius circles (as many as the points with target points based on the farthest point circle center, and its radius). In the last part of step1, the points are examined in the overlap among the points and the points out of the radius of Apollonius circle so that the points in this region are assigned to the most similar and nearest group belonging to them. For step 2, the original neighborhood groups are ready. In the example presented in Fig .4 (a)-(d) given below, out of the specified points, point 1, 9 and 11 ($T = \{1, 9, \text{and } 11\}$) have the target points indicated with red square (a), and Apollonius circles were also drawn by the farthest points {5, 10, and 13} which are shown with orange triangle. Then, Apollonius circles are drawn for forming Groups = {G1, G2, and G3}.

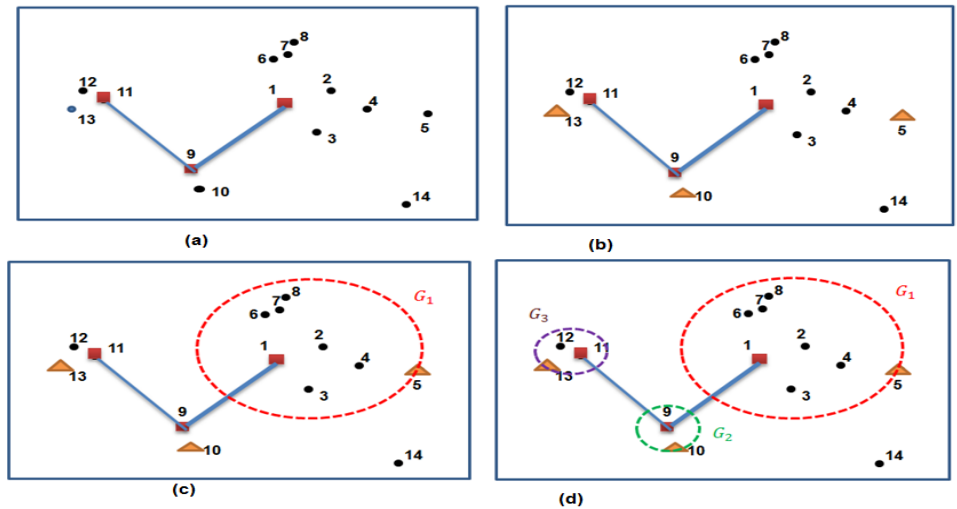

**Fig. 4.** Example data set for the NCAR algorithm, (a) points 1, 9 and 11 are target points, (b) point 5 is the farthest point of target point 1, (c) point 10 is the other farthest point of target point 9, (d) point 13 is other the farthest point of target point 11, formation G1, G2 and G3.

**Step 2: Finding the decision region inside the groups with NCARD algorithm**

In this step, finding the decision region inside the groups will be discussed. Inside the groups formed in the first step for the target points, the decision region of the points around it are examined. First, all points inside the group are arranged from the lowest distance to the highest one. Then, the Apollonius circles for every single target point and its neighboring point is drawn. Later, based on formula (2) and (3), the neighborhood regions between the target point and the points inside the main Apollonius circle will be drawn. Now if there is no other data inside the circle, the density between the target point and its neighboring point will be zero but if a point or some points are located inside the regions, the density between the target point and the respective point will be as many as the points inside the circle. For reducing calculations and not involving all points, while examining the set of data point densities after sudden density increase, suddenly we reach density of zero inside the set; this is called Control Point (CP). An example of step 2 is presented in Fig. 5 (a)-(d). The ordered set for point 1 is $T_1 = \{2, 3, 4, 6, 7, 8, \text{and } 5\}$. Point 2 is directly connected to point 1, and the nearest neighbors with indirect connection are point 3 and 4. Density values of points in $T_1 = \{2, 3, 4, 6, 7, 8, \text{and } 5\}$ are $D_1 = \{0, 1, 2, 0, 1, 2, \text{and } 3\}$ respectively. As the first density decrease occurs at point 6, it becomes a control point for point 1.

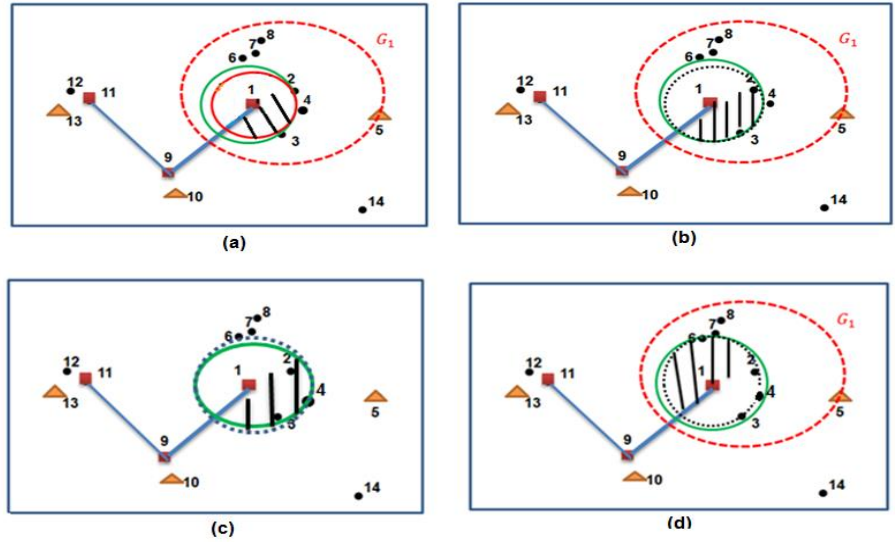

**Fig. 5.** Example data set for the NCARD algorithm, (a) decision reign 1, 2 and direct relationship, density$_{1,2}$ = 0, (b) decision reign 1, 3 and indirect relationship, density$_{1,3}$ = 1, (c) decision reign 1, 4 and indirect relationship, density$_{1,4}$ = 2, (d) decision reign 1, 6, density$_{1,6}$ = 0 and find 6 as control point.

In the next step, CP is the target point for the next analysis. The set of neighborhoods before control point for each target point are the direct and indirect neighbors of that point in which the points having zero density are direct neighbors of the target point, and the points with non-zero densities are the indirect neighbors. It should be noted that the sudden fall of density to zero indicates cluster change or change in direction since basically it is expected that by going away from the target point, the density should increase.

### Step 3: Re-clustering

The set of neighbors before each control point are the direct and indirect neighbors of that point. The neighborhood points of the control point by themselves should be compared with the neighborhood sets of the target point, and if there were any common point between the two clusters, they would be considered as indirect neighbors in Eq. (4). Otherwise, they have no relationship and are grouped in different sub-cluster (two new sub-clusters will be formed after this in Fig. 6). In this step, the outlier data are distant from the Apollonius circles and farther away from the data points inside Apollonius circles (point 5 and 14).

$$T_m \cap CP_m \neq \varnothing \qquad (4)$$

Finding the density and relationships for the target point will also continue as before. The obtained sets should have the greatest rate of intra-set and the least rate of inter-set decision region.

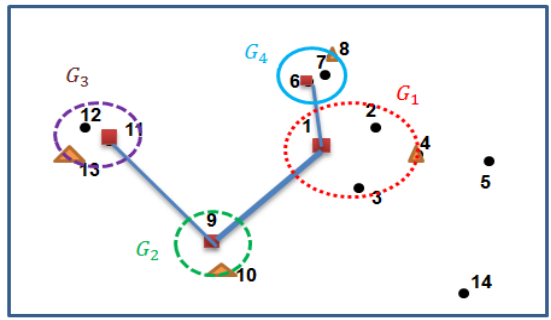

**Fig. 6.** Example for creating new sub-clusters in NCARD algorithm, point 6 is the new target point, formation G4 and point 4 is the new farthest point for point 1 so formation G1.

### 3.3. Computational complexity of NCARD algorithm

In the present paper, we analyzed the complexity of NCARD algorithm and compared it against several famous algorithms on neighborhood construction the results of which are shown in Table 3. As the table indicates, the proposed method is less complex than other geometric methods such as DPC-PCA (Du et al., 2016), DPC-KNN (Rodriguez & Laio, 2014), NCAR (Pourbahrami et al., 2018) and NC (İnkaya et al., 2015). The experiments are coded in MATLAB software on a 4 Due Intel Core, 3 GHz CPU and 4 GHz of RAM. The complexity of our algorithm is of O (n2) type; later we can use principal components analysis methods to reduce its complexity so that it would be applicable to very large data.

Table 3. Time complexity.

| NC (İnkaya et al., 2015) | DPC-PCA (Du et al., 2016) | DPC-KNN (Rodriguez & Laio, 2014) | NCAR (Pourbahrami et al., 2018) | NCARD |
|---|---|---|---|---|
| $o(n^3)$ | $o(n^3)$ | $o(n^3)$ | $o(n^2)$ | $o(n^2)$ |

### 4. Assessment Criteria

Assessing the results of algorithm clustering is highly important. In this line, several validation indexes have been proposed. In the present study, three criteria were used to assess and validate the accuracy of clustering results using NCARD algorithms; the indexes are: Rand Index (RI) and Quasi-Jaccard Index (QJI) and Jaccard Index (JI) in Eq. (5-7). These criteria are defined below (İnkaya et al., 2015, Sardana & Bhatnagar, 2014, Yu & Kim, 2016, Du & Jia, 2016, Li et al., 2018)

$$RI = \frac{a+d}{a+b+c+d} \quad (5)$$

$$JI = \frac{a}{a+b+c} \quad (6)$$

$$QJI = \frac{a+b}{a+b+c} \quad (7)$$

a: the number of pairs of observes that belong to the same target cluster and are assigned to the same closures. b: the number of pairs of observes that belong to the same target cluster but are assigned to different closures. c: the number of pairs of observes that belong to different target clusters but are assigned to the same closures. d: the number of pairs of observes that belong to different target clusters and are assigned to different closures.

RI is the most commonly used criteria for assessing clustering algorithms. This index indicates the extent of overlap resulting from clustering as well as the real set of the data.

JI is a criterion for comparing the similarities or differences among the statistical sample sets. This validation index has applications in external clustering. This index emphasizes the paired points which have been assigned to their clusters accurately.

QJI is an assessment criterion for quantifying clusters, and assessment is based on mixing data points in the clusters. In neighborhood construction algorithms, the important issue is also examining mix points in the other clusters, and the algorithms presented in this regard should be able to minimize inaccurate point mixes with other clusters. In neighborhood construction algorithms, the important issue is not distribution of target clusters, and there is no need for penalizes.

## 5. Experimental Results

In this part, the performance quality of the proposed algorithm will be examined. We will show the results of the experiments on NCARD algorithm and will indicate enhancement of its accuracy in clustering and locating outlier data in comparison with other algorithms. Neighborhood construction algorithm was compared with ε-neighborhood, KNN (Jarvis & Patric 1973, Lu & Fu 1978, Qin et al., 2018, Tao et al., 2006), NCAR (Pourbahrami et al., 2018), and NC (İnkaya et al., 2015). After normalizing the data, performance of NCARD algorithm on real and artificial data are presented. The test data sets for checking the performance of our algorithm include both real and artificial data, and the accuracy of the proposed algorithm will be separately compared against the known and the state of the art algorithms.

### 5.1 Experiments on real data sets

Table 4 shows the detailed information about the real data sets used in this paper. All real data include numerical attributes (Abualigah et al., 2017). In Fig. 7, the accuracy of our proposed method is much more than that of base density methods such as DPC-KNN (Rodriguez & Laio, 2014), DPC-PCA (Du et al., 2016) and NCAR (Pourbahrami et al., 2018), on real data. After normalizing the data, the performance of NCARD algorithm on real data has been presented. NCARD has better result than other algorithms on Pen-based digits, Sonar and Waveform. The accuracy of NCARD is better than the remaining three algorithms in high dimension data sets. NCAR is the second best, followed by DPC-KNN and DPC-PCA. In our implementation, p is the parameter which is used to define the number of neighbours in the first step process in NCARD, NCAR, DPC-KNN and DPC-PCA algorithms. In our algorithm, we select parameter p from a fixed value [5%]. For real data sets, the average run times of NCARD and NCAR are 58 and 47 seconds, respectively.

**Table 4.** Details of real-world data sets taken from UCI.

| Data set | #instance | #feature | cluster |
|---|---|---|---|
| **Iris** | 150 | 4 | 3 |
| **Wine** | 178 | 13 | 3 |
| **Heart** | 270 | 13 | 2 |
| **Waveform** | 5000 | 21 | 3 |
| **Sonar** | 208 | 60 | 2 |
| **Glass** | 214 | 10 | 6 |
| **Pen-based digits** | 109,962 | 16 | 10 |
| **LED digits** | 500 | 7 | 3 |
| **Seeds** | 210 | 7 | 3 |

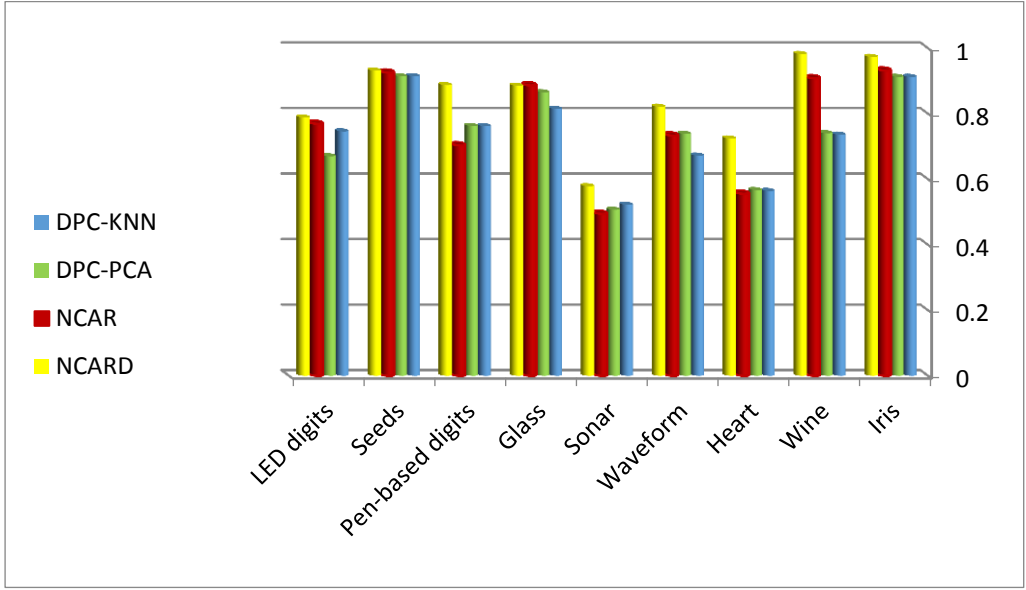

**Fig.7.** Comparison of RI values provided by different clustering methods.

### 5.2 Experiments on artificial data sets

We compared our proposed neighborhood construction algorithm with kNN, $\varepsilon$-neighborhood, NC, and NCAR. In kNN value $k$ is of 5% to 10% of points in the data set, where we donate these configurations by kNN1 and kNN2, respectively. The purpose is to make sure that the value of $k$ is at least the size of the smallest cluster. Inkaya et al. procedure is used to set the value of ε and $k$. The $k$-distance graph ($k$=4) is used to determine $\varepsilon$ value for each data set. Two groups of data sets (Group1 and Group 2) are used:

Group 1 data sets: The first group includes twenty three artificial data sets which are used in our implementation experiments in order to validate the performance of NCARD algorithm in comparison with other algorithms. Group 1 is composed of 2-dimensional and higher dimensional data sets compiled by Inkaya et al., (2015).

Group 2 data sets: The second group includes twenty four artificial data sets which are used in our implementation experiments in order to validate the performance of NCARD algorithm in comparison with other algorithms. Each data set is composed of four clusters, namely the letters S, A, O and E. The data set D-000 to D-1211 of Group 2 with 5-18 target points, 2780-2783 data points is used to illustrate the performance of the NCARD algorithm. Group 2 is composed of 2-dimensional and higher dimensional data sets compiled by Inkaya et al., (2015).

Table 5 shows the results of comparing the accuracy of NCARD clustering algorithm with NCAR, kNN, NC algorithms. According to the results in Table 5, the proposed algorithm works equally well with high and low amounts of data, and it increases the accuracy of NCAR algorithm and removes its weaknesses with high amounts of data. Our proposed method has higher level of accuracy than other geometric methods mentioned above such as NCAR, NC, as well as base distance methods such as kNN and ε-neighborhood on artificial data sets. NC, NCAR, and NCARD have significantly higher JI and RI values in comparison with KNN1, KNN2, and ε-neighborhood. The minimal NCARD performance is much higher than the other algorithms shown in Table 5. As Table 5 indicates, according to the maximum performance of all algorithms shown in the table, there is at least one algorithm in group 1 of data sets in which the target clusters would be obtained. For group 1 the min performance of NCAR and NC is not the best among the six algorithms of competing approaches, instead NCARD wins in this case. In fact, in group 1 NCARD has the best performance.

As Table 6 indicates, the separation of data sets in the clusters of group 2 is smaller than their compactness. In other words, with a decrease in the distance among the clusters, there would be more cluster mix, which is the major limitation of NCAR and NC. Furthermore, in group 2, KNN1 and KNN2 fail to locate their target clusters in the data sets of group 2. Meanwhile, NCARD, NCAR, ε-neighborhood, and NC show their best performance and find their target clusters. The smallest neighborhood variance has been reported for NCARD algorithm; this shows it leads to more homogeneous neighborhood in data sets. Furthermore, it has been reported that out of the three algorithms, NCARD, NCAR, and NC have shown rather homogeneous neighborhood. The basis for judging statistical significance of differences is the mean value of measured performance. In the two data set groups, NCARD yields significantly better performance than the other neighbourhood algorithms. NCAR, NC, and ε-neighborhood are the next best algorithms, respectively. NCARD shows better strong results in cluster mixes due to its rather low standard deviation values. JI and RI of NCARD algorithm in group 1 data set perform better than the JI and RI of NC and NCAR indicating that cluster mixes are more frequent in NCARD than NCAR. However, divided clusters are more frequent in NCARD than NCAR. The small number of cluster mixes in NCARD facilitates effective clustering. KNN1 and KNN2 have lower JI and RI values than the others.

**Table 5.** Comparison results of KNN1, KNN2, ε-neighborhood, NC, NCAR and NCARD in terms of mean accuracy, JI and QJI for Group 1 on 23 data sets.

| Accuracy and similarity | Algorithms | | | | | |
| --- | --- | --- | --- | --- | --- | --- |
|  | KNN1 | KNN2 | ε-neighborhood | NC | NCAR | NCARD |
| **RI** | | | | | | |
| Average | 0.80 | 0.75 | 0.88 | 0.91 | 0.93 | **0.96** |
| Std.dve | 0.21 | 0.22 | 0.19 | 0.10 | 0.10 | **0.11** |
| Min | 0.31 | 0.31 | 0.33 | 0.66 | 0.68 | **0.68** |
| Max | 1.00 | 1.00 | 1.00 | 1.00 | 1.00 | **1.00** |
| **JI** | | | | | | |
| Average | 0.79 | 0.74 | 0.86 | 0.88 | 0.88 | **0.89** |
| Std.dve | 0.21 | 0.22 | 0.21 | 0.13 | 0.11 | **0.08** |
| Min | 0.31 | 0.31 | 0.31 | 0.31 | 0.34 | **0.40** |
| Max | 1.00 | 1.00 | 1.00 | 1.00 | 1.00 | **1.00** |
| **QJI** | | | | | | |
| Average | 0.79 | 0.74 | 0.87 | **0.99** | **0.99** | 0.99 |
| Std.dve | 0.21 | 0.22 | 0.21 | 0.02 | **0.01** | **0.01** |
| Min | 0.31 | 0.31 | 0.31 | 0.91 | 0.90 | **0.92** |
| Max | 1.00 | 1.00 | 1.00 | 1.00 | 1.00 | **1.00** |

**Table 6.** Comparison results of KNN1, KNN2, ε-neighborhood, NC, NCAR and NCARD in terms of mean accuracy, JI and QJI for Group 2 on 24 data sets.

| Accuracy and similarity | Algorithms | | | | | |
|---|---|---|---|---|---|---|
| | KNN1 | KNN2 | ε-neighborhood | NC | NCAR | NCARD |
| **RI** | | | | | | |
| Average | 0.28 | 0.25 | 0.89 | 0.89 | 0.90 | **0.93** |
| Std.dve | 0.10 | 0.00 | 0.15 | 0.17 | 0.16 | **0.15** |
| Min | 0.24 | 0.24 | 0.51 | 0.41 | 0.45 | **0.53** |
| Max | 0.60 | 0.25 | 1.00 | 1.00 | 1.00 | **1.00** |
| **JI** | | | | | | |
| Average | 0.26 | 0.25 | 0.76 | 0.74 | 0.74 | **0.77** |
| Std.dve | 0.04 | 0.00 | 0.25 | 0.26 | 0.26 | **0.20** |
| Min | 0.24 | 0.24 | 0.33 | 0.25 | 0.30 | **0.35** |
| Max | 0.38 | 0.25 | 1.00 | 1.00 | 1.00 | **1.00** |
| **QJI** | | | | | | |
| Average | 0.26 | 0.25 | 0.82 | 0.88 | 0.89 | **0.91** |
| Std.dve | 0.024 | 0.00 | 0.24 | 0.21 | 0.20 | **0.18** |
| Min | 0.24 | 0.24 | 0.54 | 0.39 | 0.41 | **0.23** |
| Max | 0.38 | 0.25 | 1.00 | 1.00 | 1.00 | **1.00** |

# 6. Conclusion

In the present paper, we have presented an improved algorithm for neighborhood construction using Apollonius geometric decision region model and density among the points. The goal of the idea presented above was enhancing the accuracy of locating similar points and grouping them. Apollonius circle was utilized since interesting orbital decision region could be extracted for locating neighborhood among the points under study (attraction of local nearby points); it could also reduce the complexity of the previous algorithms. Moreover, the proposed algorithm is able to explore the outlier data in different noisy data sets. The experiments on real and artificial data sets indicate the high accuracy and power of NCARD algorithm for clustering high dimension data and detecting outlier data. The high efficiency of Apollonius structure in assessing the local similarities among the observations has opened a new field in the science of data mining called geometrical data mining. The proposed algorithm is more accurate than other algorithms up to almost 8-13% in real and artificial data sets. In future studies, we intend to extract the connection and density among the data points without any need to find high density points using Apollonius circle or without any need to offer a new definition for parameter. Moreover, these algorithms can be used in some areas including classification, grouping the social media, grouping the edges in visualization, and clustering wireless sensor networks.